\documentclass[3p,twocolumn]{elsarticle}

\usepackage{graphicx}
\usepackage{multirow}
\begin{document}

%\title{Strength function sum rules and the generalized Brink-Axel hypothesis\tnoteref{t1}}

\title{Systematics of strength function sum rules\tnoteref{t1}}

\author{Calvin W. Johnson}
\address{Department of Physics, San Diego State University,
5500 Campanile Drive, San Diego, CA 92182-1233}
%\affiliation{Department of Physics, San Diego State University,
%5500 Campanile Drive, San Diego, CA 92182-1233}
\ead{cjohnson@mail.sdsu.edu}
\tnotetext[t1]{This article is registered under preprint arXiv:1506.04700}

\begin{abstract}
%The expectation value is a fundamental calculational tool for quantum mechanics, including 
% many-body systems.  
 Sum rules provide useful insights into 
transition strength functions and are often  expressed as expectation values
of an operator. 
In this letter I  demonstrate that  non-energy-weighted transition sum rules
 have strong secular dependences on the 
 energy of the initial state.  Such non-trivial systematics have consequences: the 
simplification suggested by the generalized Brink-Axel hypothesis, for example, does not hold for most cases, 
though it weakly holds in at least some cases for electric dipole transitions.  
Furthermore, I show the systematics can be understood through 
spectral distribution theory, calculated via traces of operators and of products of operators. 
Seen through this lens, violation of the generalized Brink-Axel hypothesis is unsurprising: 
one \textit{expects} sum rules to evolve with excitation energy. Furthermore, to lowest order the slope 
of the secular evolution can be traced to a component of the Hamiltonian being positive (repulsive) 
or negative (attractive). 
%This adds to the body of evidence that the generalized Brink-Axel hypothesis does not hold.
\end{abstract}


\maketitle

%Among of the first skills physicists learn in a quantum mechanics course is how to compute, for a given state, the expectation value of
%an observable represented by an operator.  

%Expectation values, representing the average of many measurements, are easier to compute than 
%the probabilities of individual outcomes. 
 % While expectation values are 
%most often calculated for the ground state, one can of course calculate them for excited states as well.  We seldom 
%think about the relationship between expectation values for different states, however. 

%Because of the relative ease of computing expectation values, we value them in summarizing properties of quantum systems, 
%such as sum rules for transition matrix elements, also called strength functions,

One important way to investigate quantum systems both experimentally and theoretically are through  strength functions,
\begin{equation}
S(E_i,E_x) = \sum_f \delta(E_x - E_f + E_i) | \langle f | \hat{\cal O} | i \rangle |^2.
\label{strengthfunction}
\end{equation}
which is the probability to transition from a state at initial energy $E_i$ to some final state 
 at an energy $E_f =E_i + E_x$, via  the operator $\hat{\cal O}$; $| i \rangle$ and $| f\rangle$ 
 are initial and final states, respectively.
%I allow for excited initial states for applications to, say, the gamma-decay cascade of a highly excited nucleus, 
%as well as in hot, dense environments 
%such as those found in astrophysics. 
In particular I consider strength functions sum rules for atomic nuclei for transitions 
%In this letter I will first discuss nuclear strength functions, the widely-used and 
%simplifying assumption of the Brink-Axel hypothesis (broadly speaking, the hypothesis that the strength function is independent of the initial state), 
%and sum rules for strength functions which can expressed as 
%expectation values.   Numerical calculations for a variety of transition operators and nuclides 
%illustrate that the non-energy weighted sum rule shows simple behavior, but not as 
%simple as  Brink-Axel suggests. I can explain the evolution of sum rules--or, indeed, the expectation value of any operator--with excitation energy, including fluctuations, 
%using spectral distribution theory, which relies upon traces of operators and 
%products of operators.  This will allow us to understand the conditions under which the Brink-Axel hypothesis will hold, and when it will be violated. Furthermore, 
%although I focus on applications to nuclear transition strength functions, the framework holds broadly for any quantum mechanical system.
 such as electric dipole (E1), magnetic dipole (M1), electric quadrupole (E2), Gamow-Teller (GT), and others \cite{BM98,Lawson80}. % (though
% the general principles apply more generally).  
%These transitions govern decays,  capture reactions, and inelastic scattering.
%In  interactions with photons and leptons the Born approximation is often sufficient, so that one only needs 
%the single transition probability $|\langle f | \hat{\cal O} | i \rangle|^2$, where $i$ and $f$ label 
%initial and final eigenstates of the nuclear Hamiltonian and $\hat{\cal O}$ is the appropriate 
%transition operator. 
 Such transitions not only provide important diagnostics of nuclear structure, and thus test 
 theoretical descriptions of nuclei against experiment, but also have important impacts in 
 astrophysical physical processes such as nucleosynthesis \cite{FFN80,FM95,LG10,LB13,TGLS15},  neutrino transport \cite{MFB14}, in the experimental extraction of the density of states \cite{SBGM00,SABG03,VGAA06,SGIL09}, and so on. 
These applications include initial states excited far above the ground state.

%A transition operator  typically couples the initial state to many final states, usually expressed 
 %as a \textit{strength function}, 
%\begin{equation}
%S(E_i,E_x) = \sum_f \delta(E_x - E_f + E_i) | \langle f | \hat{\cal O} | i \rangle |^2.
%\label{strengthfunction}
%\end{equation}
%which is the probability to transition from a state at initial energy $E_i$ to some final state 
%separated at an energy $E_f =E_i + E_x$.  I allow for excited initial states for applications to, say, the gamma-decay cascade of a highly excited nucleus, 
%as well as in hot, dense environments 
%such as those found in astrophysics. 

Often one sees the strength function displaying either a sharp or a broad peak, which 
is called a resonance, and if most of the strength is in that peak, it is a \textit{giant} resonance \cite{ring}. 
Giant resonances can have an intuitive picture: for example, for the giant (electric) dipole resonance, 
or GDR, one envisions protons and neutrons collectively oscillating against each other \cite{GT48,SJ50}. 

The Brink-Axel hypothesis \cite{Br55,Ax62} states that if the ground state has a giant electric dipole 
resonance, then 
the excited states should have giant dipole resonances as well; because the GDR is explained by 
a collective proton-versus-neutrons oscillations, it should not be very sensitive to the details of 
the initial state, and so the strength function  $S(E_i, E_x)$ in (\ref{strengthfunction}) should
independent, or nearly so, of $E_i$ \footnote{For some 
 historical details, see http://www.mpipks-dresden.mpg.de/~ccm08 /Abstract /Brink.pdf and 
http://tid.uio.no/ workshop09 /talks /Brink.pdf}. 
 While the original Brink-Axel hypothesis only
concerned  the GDR, it later became a simplifying assumption applied to more general transitions, for example M1 and GT.
As strength functions off excited state are 
particularly difficult to measure experimentally, this hypothesis, if true, 
would be  very useful.
%relevant to experimental extraction of level densites \cite{SBGM00,SABG03,VGAA06,SGIL09}.
%and astrophysics \cite{FFN80,FM95,LG10,LB13,MFB14,TGLS15}.

But is the Brink-Axel  hypothesis  true, especially for transitions other than electric dipole? 
And if it is not true, can we do anything 
about it?

Despite wide usage and some early experiments in support of the Brink-Axel hypothesis \cite{RSS81}, there is considerable evidence the Brink-Axel hypothesis fails or is 
modified for E1 \cite{RBK93,BCM95,LGCI07,AHK12,LB13}, M1 \cite{LGCI07,SFL13,BL14},
 E2 \cite{LGCI07,Sch14}, and Gamow-Teller \cite{FB97,NS07,MFB14}  transitions. 
 Nonetheless, as stated in a recent Letter \cite{SFL13}, ``It is quite common to adopt the
so-called Brink-Axel hypothesis which states that the
strength function does not depend on the excitation energy.''
  On the other hand, a recent \textit{ab initio} calculation supported the 
 Brink-Axel hypothesis for E1 transitions from low-lying states \cite{KOJ15} and the success and consistency of the Oslo method for determining the level density relies upon M1 strengths 
 following the Brink-Axel hypothesis \cite{SBGM00,SABG03,SGIL09,VGAA06}.

\begin{figure}
\centering
\includegraphics[scale=0.35,clip]{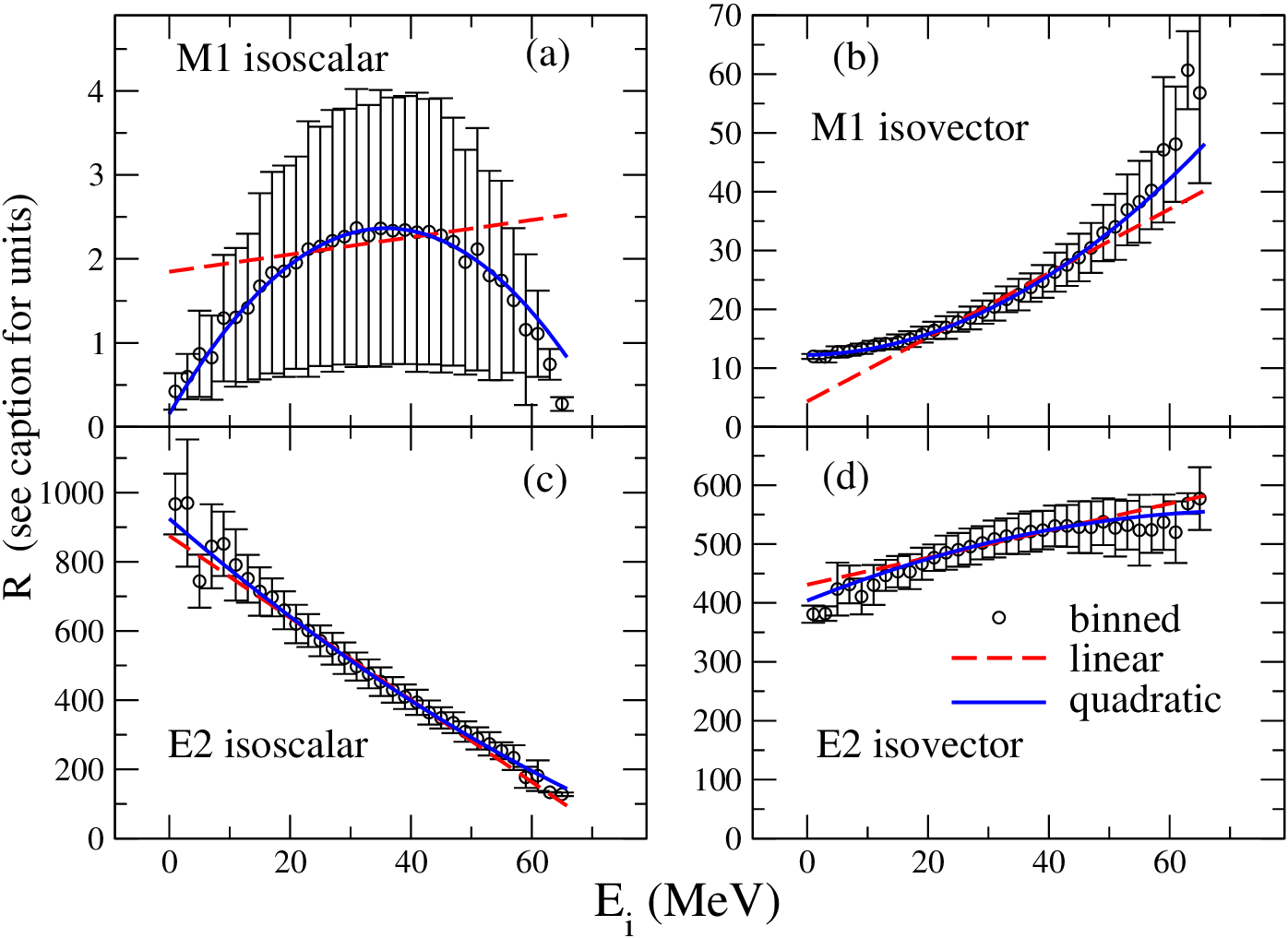}
\caption{ %(Color online) 
Non-energy weighted sum rules (here denoted as $R$) as a function of initial energy $E_i$ 
for $^{33}$P computed in the $sd$ space.  The units for summed M1 strengths (plots (a) and (b)) are $\mu_N^2$ ($\mu_N=$ nuclear magneton), while those for
E2 (plots (c) and (d)) are $e^2$-fm$^4$. The (red) dashed line is the linear approximation to the secular behavior of $R$, 
derived from spectral distribution theory, while the (blue) solid line is the quadratic approximation. }
\label{P33allstrengths}
\end{figure}

To simplify the question, I  focus on sum rules derived from  (\ref{strengthfunction}),  in particular the \textit{total strength} 
or \textit{non-energy-weighted sum rule} (NEWSR),
\begin{eqnarray}
R(E_i) = \int S(E_i,E) dE = \sum_f | \langle f | \hat{\cal O} | i \rangle |^2 \nonumber \\
= \langle i | \hat{\cal O}^\dagger \hat{\cal O} | i \rangle = \langle i | \hat{R} | i \rangle,
\end{eqnarray}
where for convenience I've defined $\hat{\cal O}^\dagger \hat{\cal O}  = \hat{R}$.  If $\hat{\cal O}$ is a 
non-scalar operator with angular momentum rank $K$ and isospin rank $I$, then
$\hat{R} = \sum_{M,\mu} (-1)^{M+\mu}$ $ \hat{\cal O}_{K \, -M, I \, -\mu} \hat{\cal O}_{K \, M, I \, \mu}$. (This definition 
includes a sum over charge-changing transitions for $I=1$, but in return $\hat{R}$ is a simpler, isoscalar operator; I 
have no reason to believe this qualitatively changes any of my conclusions.) 
Many other sum rules, such as the energy-weighted sum rule (EWSR), can also be written as expectation values of 
operators \cite{ring}, although here I will only consider the NEWSR.

If the Brink-Axel hypothesis were  true, then $R(E_i)$ would be a constant. % independent of $E_i$.
 I investigate the systematics of the non-energy-weighted sum rule  
for several different operators $\hat{\cal O}$ and nuclides as a function of the initial energy $E_i$. 
% I will show that (a) the sum rule often has a strong secular dependence upon $E_i$, 
%contrary to the Brink-Axel hypothesis, both from (b) specific microscopic nuclear calculations 
%carried out in the context of 
%the configuration-interaction (CI) shell model, as well as from (c) more general mathematical 
%considerations. I will then use the latter to give a simple parameterization of the 
%secular dependence of $R(E_i)$ using quantities relatively easy to calculate. 
To test whether or not $R(E_i)$ does or does not vary with initial energy $E_i$, I first carry out 
calculations in a detailed microscopic model, the configuration-interaction (CI) shell model.
In the CI shell model, one diagonalizes the many-body Hamiltonian in a finite-dimensioned, orthonormal basis 
of Slater determinants, which are antisymmeterized products of single-particle wavefunctions, typically 
expressed in an occupation representation \cite{BG77}. The advantage of CI shell model 
calculations is that one can generate excited states easily, and for a modest dimensionality 
one can generate all the eigenstates in the model space. 

I use the {\tt BIGSTICK} CI shell model code \cite{BIGSTICK}, which calculates the 
many-body matrix elements $H_{\alpha \beta} = \langle \alpha | \hat{H} | \beta \rangle$ and then solves 
%\begin{equation}
$\hat{H} | i \rangle = E_i | i \rangle.$
%\end{equation}
Greek letters ($\alpha, \beta, \ldots$) enote generic  basis states,
while lowercase Latin letters ($i,j,\ldots$)  label eigenstates. As {\tt BIGSTICK} computes not 
only the energies but also the wavefunctions, the sum rule $R(E_i)$ is  an expectation value 
easy to calculate.

For this study I use phenomenological spaces and interactions, although one could also 
consider \textit{ab initio} calculations as well; the latter tend to have very large dimensions though, 
making them less practical for studying the secular behavior over many MeV.  Instead I carried out 
calculations in the   $1s_{1/2}$-$0d_{3/2}$-$0d_{5/2}$ or $sd$ shell, %(sometimes 
%called the $sd$-shell, as it comprises a major harmonic oscilator shell), 
using a universal $sd$ interaction version 
`B' (USDB)~\cite{br06}. I also consider the following transition operators: M1, E2, and Gamow-Teller using
their standard forms \cite{BM98,ring,BG77}.
I do not use effective charges, I use harmonic oscillator wavefunctions with an oscillator length of 2.5 fm,  I divide sum rules for isovector operators by 3 to roughly average 
over charge-changing transitions, and use a quenched value of $g_A \approx 1 $; these assumptions are scaling factors and do not affect my conclusions.

\begin{figure}
\centering
\includegraphics[scale=0.35,clip]{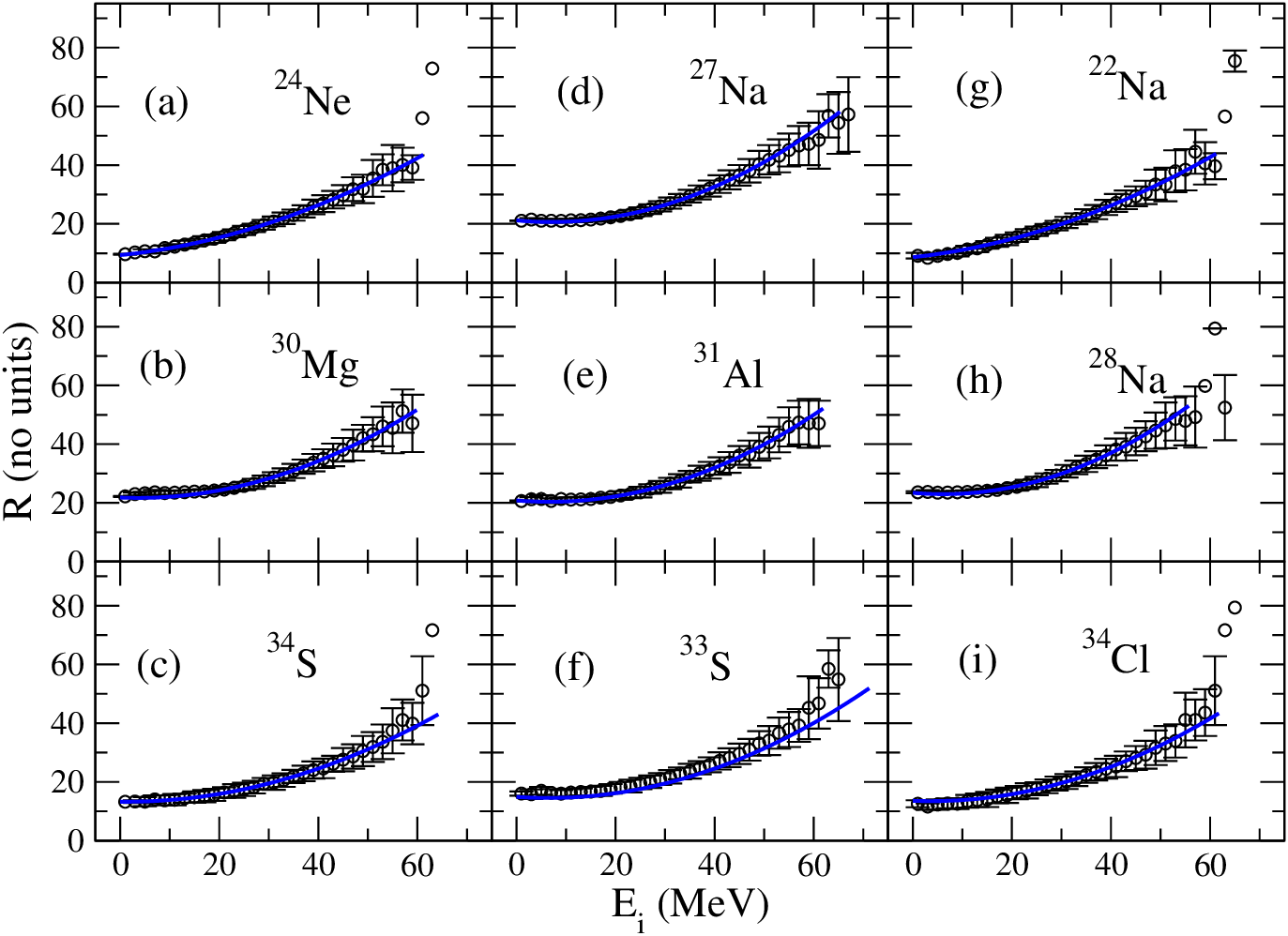}
\caption{ %(Color online) 
Non-energy weighted  Gamow-Teller sum rules (here denoted as $R$) as a function of initial energy $E_i$ 
for several nuclides in the $sd$ space.  The (blue) solid line is the quadratic approximation to the secular behavior of $R$, 
derived from spectral distribution theory. }
\label{GT}
\end{figure}

Fig.~\ref{P33allstrengths} shows the NEWSR as a function of initial energy (relative to the ground state) for the nuclide $^{33}$P for isoscalar and isovector M1 and E2 
transitions, while Fig.~\ref{GT} shows the NEWSR for Gamow-Teller for several even-even, odd-odd, and odd-$A$ nuclides. (Because I am taking the sum over 
charge-changing transitions and not the difference, the Ikeda sum rule will not tumble out of these calculations.) I binned the NEWSR into 2 MeV bins, but found the 
size of the fluctuations shown by errors bars to be insensitive to the size of the bins.  Other calculations not shown 
show qualitatively similar results, which can be summarized as: 

\noindent $\bullet$ Both the secular (average) behavior of the NEWSR and fluctuations thereof show surprisingly smooth behavior. 

\noindent $\bullet$ As illustrated for Gamow-Teller transitions in  Fig.~\ref{GT} (and duplicated but not shown for other operators), the behavior is relatively insensitive to the nuclide.

\noindent $\bullet$ Nonetheless, the behavior does depend sharply upon the transition: isoscalar E2 falls sharply with initial energy, isovector M1 and Gamow-Teller grow, % with initial energy, 
and isoscalar M1 has large fluctuations.

As the Brink-Axel hypothesis predicts NEWSR independent of $E_i$, these results numerically confirm the previously mentioned experimental and theoretical evidence against 
Brink-Axel.  
%(One immediately obvious counter-hypothesis is that the finite model space may ``cause'' the violation of Brink-Axel. This seems unlikely. The Gamow-Teller calculations, 
%for one, are computed in a complete space; the Gamow-Teller operator cannot take one out of the $sd$-space. Furthermore, if one argues that a finite space undercuts the violation of 
%Brink-Axel, a corollary would be that \textit{any} calculation of transitions must be similarly flawed.) 
Given the simple yet non-trivial systematics,  can one understand these results from basic principles?  

 If one is computing $R(E_i)$ in a finite model space, such as 
the CI shell model, and that model space has dimension $N$, then one can compute 
the \textit{average} sum rule, that is, the average expectation value, 
\begin{equation}
 \frac{1}{N} \sum_{i=1}^N \langle i | \hat{\cal O}^\dagger \hat{\cal O} | i \rangle 
= \frac{1}{N} \sum_i \langle i | \hat{R} | i \rangle  \equiv \overline{\langle \hat{R} \rangle},
%\equiv \overline{\langle \hat{R} \rangle}. 
\label{trace}
\end{equation}

If we compute the matrix elements of $\hat{R}$ in some orthonormal many-body basis, 
for example Slater determinants in the framework of the CI shell model, the sum is 
just a trace of the matrix. 
Because a trace is invariant under a unitary transformation, we can sum over any 
convenient set of basis states $\{ | \alpha \rangle \}$. %, so that 
%\begin{equation}
%\overline{R} = \frac{1}{N} \sum_\alpha \langle \alpha | \hat{R} |\alpha \rangle .
%\end{equation}
This invariance under the trace is important because the trace can be used as an inner product in 
the space of Hermitian operators in 
the framework of  \textit{spectral distribution theory} (SDT), also sometimes called
\textit{statistical spectroscopy} \cite{BF66,Fre67,Fre71,Mon75,Fre83,Won86,LSD14}.  

%But before developing this idea further, consider that 
[The notation $\langle \hat{R} \rangle 
= \langle i | \hat{R} | i \rangle$ signifies  the expectation value being an average \textit{over many measurements 
for the same state}.   Yet for SDT one averages the expectation value over all states in a 
space, usually defined by fixed quantum numbers such as the number of particles.  Practitioners 
of SDT frequently use the notation $\langle \hat{A} \rangle^{(m)}$ for (\ref{trace}) \cite{Fre83,Won86,LSD14}, where $m$ 
denotes the number of particles and possibly other quantum numbers, and the trace is implied to 
be restricted to states with those quantum numbers.  Because of the unfortunate possibility of 
confusion with the expectation value proper, I introduced a hybrid notation $\overline{\langle \hat{R} \rangle}$ 
for the average (\ref{trace}).]

To see if the sum rule $R(E_i)$ does indeed have a secular dependence 
upon the initial energy $E_i$, one can take a weighted average, namely, 
\begin{equation}
  \frac{1}{N} \sum_i R(E_i) E_i =   \frac{1}{N} \sum_i \langle i | \hat{R} \hat{H} | i \rangle  = \overline{ \langle \hat{R}\hat{ H}\rangle } .
\end{equation}
Again, since this is a trace, one can compute in any convenient basis. %, so that 
%the average is written
%\begin{equation}
%\overline{ R H }  = \frac{1}{N} \sum_\alpha \langle \alpha | \hat{R} \hat{H} |\alpha \rangle .
%\end{equation}
%In the development of spectral distribution theory, 
French proposed \cite{Fre67} the following inner product between two Hamiltonians, or more 
broadly between two Hamiltonian-like (Hermitian and angular momentum scalar) operators:
\begin{eqnarray}
\left (\hat{H}_1, \hat{H}_2  \right) = 
\overline{
 \left \langle \left (  \hat{H}_1 -\overline{  \langle \hat{H}_1 }\rangle  \right) 
\left (  \hat{H}_2- \overline{ \langle \hat{H}_2 \rangle} \right) \right \rangle } \nonumber \\
=  \overline{ \langle \hat{H}_1 \hat{H}_2 \rangle} - \overline{\langle \hat{H}_1 \rangle} \,\,\,
 \overline{\langle \hat{H}_2\rangle}.
\end{eqnarray}
%where, to be clear
%\begin{equation}
%\overline{ A} = \frac{1}{N} \sum_{\alpha \in {\cal S}} \langle \alpha| \hat{A} | \alpha \rangle
%\end{equation}
%where ${\cal S}$ is some many-body space and $N$ is the dimension of that space. 
The appeal of this definition of the inner product between Hamiltonian-like operators is that, 
if the operators are angular momentum scalars and if one works in a finite, spherically symmetric 
shell-model single-particle space, one can calculate the 
traces directly without constructing the matrix \cite{BF66,Fre83,Won86,LSD14}. One can  sum over 
states with specified isospin (while one can take sums over specified angular momentum \cite{SHZ13}, the 
resulting formulas are significantly more tedious and computationally intensive) or even just on 
subspaces defined by configurations, that is, a fixed number of particles in each orbit.  In principle one can 
 take higher-order moments or work with Hamiltonians or particle rank higher than two.
For this work, however, I use a recent code \cite{LSD14} which reads in only isospin-invariant two-body 
interactions and which calculates at most second moments (i.e., the inner product defined above) 
working in spaces with fixed total number of valence particles $A_\mathrm{val}$
and total isospin $T$.

With the definition of an inner product in the space of operators, we can return to the 
question of the invariance of the strength function with initial energy. One necessary, but 
by no means sufficient, condition 
for the invariance of the strength function is that the total strength not change with 
initial energy, that is, $R(E_i) \approx $ constant. Such a condition implies 
$
\overline{ \langle \hat{R}\hat{H}\rangle } \approx \overline{\langle \hat{R}\rangle } \,\,\,  \overline{ \langle \hat{H}\rangle },
$
but this reduces to the inner product $( \hat{R}, \hat{H} ) = \overline{ \langle \hat{R}\hat{H}\rangle } - \overline{\langle \hat{R}\rangle } \,\,  \overline{\langle\hat{H}\rangle } \approx 0$, that is, 
the Hamiltonian $\hat{H}$ and the operator $\hat{R}$ must be ``orthogonal'' in a well-defined way.  While this could happen 
by accident, in general it will not, as we already see in the examples above.

 As it turns out, the above condition corresponds to the linear dependence of $R(E_i)$ on $E_i$. We can go to 
 a higher order polynomial description, especially if we assume that the state density  of the many-body Hamiltonian 
is well-described by a Gaussian  with centroid $\bar{E}$ and width $\sigma$, that is, 
\begin{equation}
\rho(E) = N (2\pi \sigma^2)^{-1/2} \exp \left( - \frac{(E-\bar{E})^2}{2\sigma^2} \right ),
\end{equation}
which is often a good assumption for nuclei \cite{Mon75}. 
In the language of spectral distribution theory,
\begin{eqnarray}
\bar{E} = \overline{\langle \hat{H}\rangle },\\
\sigma^2 = \overline{\langle \hat{H}^2\rangle} - \left( \overline{H} \right)^2.
\end{eqnarray}
Let's further assume that the sum rule 
$R(E)$ is a quadratic polynomial in $E$:
\begin{equation}
R(E) = R_0 + R_1 \frac{ ( E - \bar{E}) }{\sigma} 
+ R_2 \frac{ ( E - \bar{E})^2 }{\sigma^2}.  
% + R_3 \frac{  ( E - \bar{E})^3 }{\sigma^3}  .
\end{equation}
Then one can easily compute the following averages:
\begin{eqnarray}
 \overline{R}= \overline{\langle \hat{R}\rangle } = R_0 + R_2;   \label{Rpoly0}\\
( \hat{R}, \hat{H})/ \sigma = R_1. \label{Rpoly1} % + 3R_3.
\end{eqnarray}
One could add an additional constraint by by higher moments, for example $\overline{\langle \hat{R}\hat{H}^2\rangle}$. While such 
higher moments are calculable \cite{Won86}, the formula are cumbersome and prone to slow evaluation; furthermore experience in 
unpublished work suggests even higher moments have difficulty in describing the tails of distributions \cite{JNO01}. (This is understandable; the traces 
are just averages, after all, and dominated by the density of states in the middle of the spectrum.)
Instead, I  use the sum rule at the ground state energy $E_{gs}$, which is often accessible:
\begin{equation}
R(E_{gs}) = R_0 +  R_1 \frac{ ( E_{gs} - \bar{E}) }{\sigma} 
+ R_2 \frac{  (E_{gs}  - \bar{E})^2 }{\sigma^2}  \label{Rpoly2}. 
% + R_3 \frac{  ( E_{gs}  - \bar{E})^3 }{\sigma^3}.
\end{equation}
These  equations (\ref{Rpoly0},\ref{Rpoly1},\ref{Rpoly2}) can be easily solved.

Fig~\ref{P33allstrengths} shows both linear (red dashed lines) and quadratic (blue solid lines) approximations to $R(E_i)$.  Although the linear 
approximation demonstrates a secular dependence on $E_i$, in general the quadratic does better in describing the 
secular evolution of the sum rule. Fig.~\ref{GT} shows only the quadratic approximation.

%[Discussion of fluctuations]

Now, as illustrated in the figures, while one has smooth secular behavior, there are nontrivial fluctuations about the secular trends. 
The fluctuations are insensitive to the size of the energy bins. Although the fluctuations about the smooth secular behavior are not 
easily written in terms of traces,  one might be able to derive the fluctuations from random matrix theory; but this will have to 
be left to future work.

%\bigskip

%[Final Illustration with pairing interaction ]

%Although I have been motivated by, and presented examples for, the NEWSR for transitions, these results apply to any operator.  For example, in Fig.~TBA, I show how 
%the expectation value of the pairing Hamiltonian evolves with the excitation energy of the state.

\begin{figure}
\centering
\includegraphics[scale=0.35,clip]{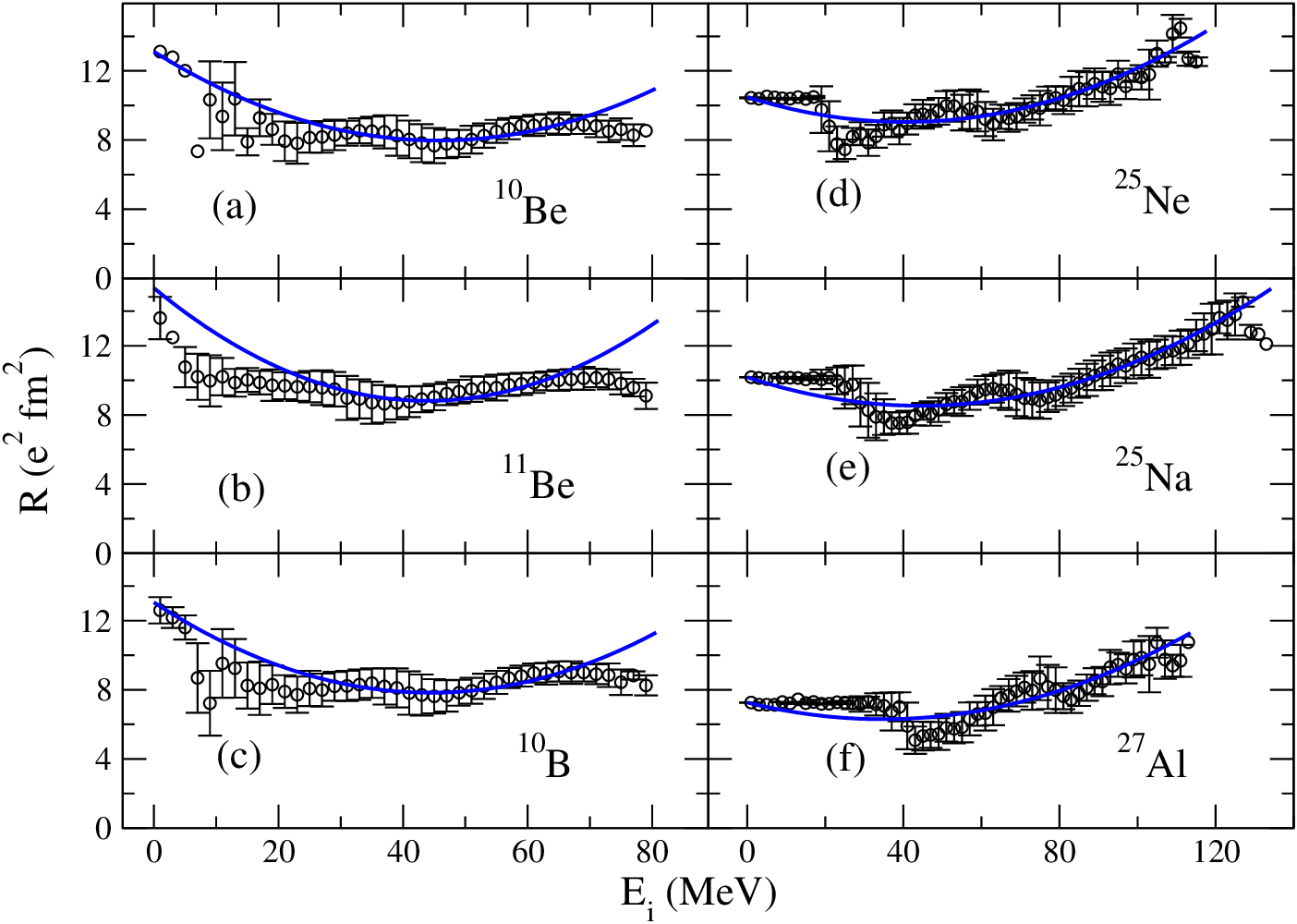}
\caption{ %(Color online) 
Non-energy weighted isovector E1 sum rules (here denoted as $R$) as a function of initial energy $E_i$ 
for several nuclides in the $p$-$sd_{5/2}$ space.   The (blue) solid line is the quadratic approximation to the secular behavior of $R$, 
derived from spectral distribution theory.}
\label{E1}
\end{figure}

The original Brink-Axel hypothesis described E1 strength functions. To  explore them, I use a space with opposite parity orbits, the
$0p_{1/2}$-$0p_{3/2}$-$1s_{1/2}$-$0d_{5/2}$ or $p$-$sd_{5/2}$ space, chosen so I could fully diagonalize for some nontrivial cases. For an interaction I
 use the Cohen-Kurath (CK) matrix elements in
the $0p$ shell\cite{CK65}, the older USD interaction \cite{Wildenthal} in
the $0d_{5/2}$-$1s_{1/2}$ space, and the Millener-Kurath (MK)
$p$-$sd$ cross-shell matrix elements\cite{MK75}.  Within the $p$ and
$sd$ spaces I use the original spacing of the single-particle
energies for the CK and USD interactions, respectively, but then
shift the $sd$ single-particle energies up or down relative to the
$p$-shell single particle energies to get the first $3^-$ state at approximately
$6.1$ MeV above the ground state. The rest of the spectrum, in
particular the first excited $0^+$ state, is not very good, but the
idea is to have a non-trivial model, not exact reproduction of the
spectrum.  Because this space does not allow for exact center-of-mass projection I restrict myself to isovector E1 transitions. The resulting 
NEWSRs are shown in Fig.~\ref{E1}, illustrating only a weak violation of Brink-Axel.

Although the quadratic approximation captures the general trends, the secular behavior for $R(E)$ is not as smooth. This may be because of the model space. 
The density of states for these nuclides, for example, are not as Gaussian-like as for the $sd$-shell examples shown; the beryllium and boron nuclides have large
third moments, while the the neon, sodium, and aluminum nuclides have larger fourth moments (``fat tails'') than Gaussians. 

Nonetheless, not only do we have evidence that the generalized Brink-Axel hypothesis is not followed, we can understand 
\textit{why}. Previous work has suggested specific reasons for breaking the Brink-Axel hypothesis: changes in deformation as one goes up in energy 
explains the increase in width for the GDR \cite{BCM95}, while a decrease in spatial symmetry/increase in SU(4) 
symmetry explains the increase of strength in Gamow-Teller sum rules \cite{FB97}. Spectral distribution theory provides  a 
more general understanding.  By establishing a vector space for Hamiltonians  such that 
\begin{equation}
\hat{H} = \sum_\sigma  c_\sigma \hat{R}_\sigma,
\end{equation}
the inner product defined by SDT yields 
$(\hat{H}, \hat{R}_\sigma) = c_\sigma $ (up to some easily-defined normalization). Here is the key point: if $c_\sigma < 0$, that is, attractive, 
one expects a negative slope to $R(E)$ and more strength for low-lying initial states. This is seen in Fig~(\ref{P33allstrengths})(c),where the 
operator $\hat{R} \sim Q \cdot Q$, the  quadrupole-quadrupole interaction.  If, on the other hand, $c_\sigma > 0$, that is repulsive, as 
for $(\sigma \tau)^2$ as in Fig.~(\ref{GT}), low-lying states have less total strength. Only if $c_\sigma \approx 0$ could the Brink-Axel hypothesis 
be true, at least at the lowest level. 
Of course, the linear approximation is not  always sufficient to fully describe the secular behavior; for many cases one needs at least quadratic and possibly even higher-order 
terms \cite{JNO01}.
% Nonetheless, one can make qualitative, if not quantitative predictions, even for operators beyond the usual electromagnetic and weak transitions. For 
%example, nuclear forces are well-known to have an attractive pairing component, $\hat{P}^\dagger \hat{P}$ where  $\hat{P}^\dagger$ is the pair creation operator. Then my basic argument 
%suggests that $\langle i |  \hat{P^\dagger} \hat{P} | i \rangle$ should be largest for low-lying states, and indeed I can demonstrate this is the case for several $sd$-shell nuclides 
%in Fig.~4. [To be inserted]  This result is more commonly expressed as low-lying eigenstates being dominated by low-seniority states, but this is just another way of stating my 
%general result.

%\begin{figure}
%\centering
%\includegraphics[scale=0.45,clip]{sdpair}
%\caption{(Color online)  Expectation value of the pairing Hamiltonian $\langle \hat{P}^\dagger \hat{P} \rangle$ as a function of 
%energy $E_i$. The (blue) solid line is the quadratic approximation to the secular behavior, 
%derived from spectral distribution theory, while the (red) dashed line is the linear approximation.}
%\label{pairing}
%\end{figure}

In summary, I have numerically demonstrated that that the  non-energy weighted sum rule for  transition operators applied to several sample  nuclides 
evolves  with the energy of the initial state--weakly for the case of isovector E1, and more strongly for other operators--and furthermore that 
such variation is \textit{expected} from spectral distribution theory. In particular, one can predict qualitatively whether a sum rule will grow or shrink in magnitude 
with initial energy, depending if part of the Hamiltonian (that part proportional to the operator $\hat{R}$ for the sum rule, that is, the square of 
the transition operator for the NEWSR) is attractive or repulsive.  
In many 
cases one needs higher moments for accurate quantitative predictions , but it should be clear now that 
 one should only invoke Brink-Axel  with  caution.  
%There is hope, however, that one could use spectral 
%distribution theory to determine cases with a strong energy dependence, although computing sufficiently high moments for accuracy is not a simple thing.

This material is based upon work supported by the U.S. Department of Energy, Office of Science, Office of Nuclear Physics, 
under Award Number  DE-FG02-96ER40985.


\begin{thebibliography}{99}

\bibitem{BM98} A.~ Bohr and B.~R.~Mottelson, \textit{Nuclear structure}, (World Scientific, Singapore, 1998).

\bibitem{Lawson80} R. ~D.~Lawson, \textit{Theory of the nuclear shell model}, (Clarendon Press, Oxford, 1980).


\bibitem{FFN80} G. M. Fuller, W. A. Fowler, and M. J. Newman, Astrophys. J. (Supplement) \textbf{42} (1980) 447.

\bibitem{FM95} G. M. Fuller and B. S. Meyer, Astrophys. J. (Supplement) \textbf{453} (1995) 792.

\bibitem{LG10} A. C. Larsen and S. Goriely, Phys. Rev. C \textbf{82} (2010) 014318.

\bibitem{LB13} E. Litvinova and N. Belov
Phys. Rev. C \textbf{88}  (2013) 031302(R).

\bibitem{TGLS15} N. Tsoneva, S. Goriely, H. Lenske, and R. Schwengner
Phys. Rev. C \textbf{91} (2015) 044318. 


\bibitem{MFB14}  G. W. Misch, G. M. Fuller, and B. A. Brown, Phys. Rev. C \textbf{90} (2014) 065808. 

\bibitem{SBGM00} 
A. Schiller, L. Bergholt, M. Guttormsen, E. Melby,  \textit{et al.} % J. Rekstad, and S. Siem, 
Nucl. Instrum. Methods Phys. Res. A \textbf{447} (2000) 498.

\bibitem{SABG03}
A. Schiller, E. Algin, L. A. Bernstein, P. E. Garrett, \textit{et al.}, %M. Guttormsen, M. Hjorth-Jensen, C. W. Johnson, G. E. Mitchell, J. Rekstad, S. Siem, A. Voinov, and W. Younes
Phys. Rev. C \textbf{68} (2003) 054326.

\bibitem{SGIL09}
N. U. H. Syed, M. Guttormsen, F. Ingebretsen, A. C. Larsen, \textit{et al.}, % T. Lönnroth, J. Rekstad, A. Schiller, S. Siem, and A. Voinov
Phys. Rev. C \textbf{79} (2009) 024316.


\bibitem{VGAA06}
A. V. Voinov, S. M. Grimes, U. Agvaanluvsan, E. Algin, \textit{et al.}, Phys. Rev. C  \textbf{74} (2006) 014314.

\bibitem{ring}
P. Ring and P. Shuck, \textit{The nuclear many-body problem} (Springer-Verlag, New York 1980).




\bibitem{GT48}  M. Goldhaber and E. Teller, Phys. Rev.  \textbf{74} (1948) 1046.

\bibitem{SJ50} H. Steinwedel and J. H. D. Jensen, Z. Naturforsch. \textbf{A5} (1950) 413. 

\bibitem{Br55} D. Brink, D. Phil. thesis, Oxford University (unpublished), 1955.

\bibitem{Ax62} P. Axel, Phys. Rev. \textbf{126} (1962) 671.

%\bibitem{Br08} http://www.mpipks-dresden.mpg.de/~ccm08 /Abstract /Brink.pdf; 
%http://tid.uio.no/ workshop09 /talks /Brink.pdf



\bibitem{RSS81}  S. Raman, O. Shahal, and G. G. Slaughter, Phys. Rev. C \textbf{23} (1981) 2794.


\bibitem{RBK93} J. Ritman, F.-D. Berg, W.~K\"uhn, V.~Metag, \textit{et al}, Phys. Rev. Lett. \textbf{70} (1993) 533.

\bibitem{BCM95} A. Bracco, F. Camera, M. Mattiuzzi, B. Million, \textit{et al.}, Phys. Rev. Lett. \textbf{74} (1995) 3748.


\bibitem{LGCI07} A. C. Larsen, M. Guttormsen, R. Chankova, F. Ingebretsen, \textit{et al.}, %T. Lönnroth, S. Messelt, J. Rekstad, A. Schiller, S. Siem, N. U. H. Syed, and A. Voinov
Phys. Rev. C \textbf{76} (2007) 044303.

\bibitem{AHK12} C. T. Angell, S. L. Hammond, H. J. Karwowski, J. H. Kelley, \textit{et al.}, Phys. Rev. C. \textbf{86} (2012) 051302(R).

\bibitem{BL14} B. Alex Brown and A. C. Larsen, Phys. Rev. Lett. \textbf{113} (2014) 252502.

\bibitem{SFL13} R. Schwengner, S. Frauendorf, and A. C. Larsen, Phys. Rev. Lett.  \textbf{111} (2013) 232504 . % arXiv:1310.7667

\bibitem{Sch14} R. Schwengner, Phys. Rev. C \textbf{90} (2014)  064321. %arXiv:1411.0876


\bibitem{FB97}  N. Frazier, B. A. Brown, D. J. Millener, and V. Zelevinsky, Phys. Lett. \textbf{B 414}(1997) 7.

\bibitem{NS07} J.-U. Nabi and M. Sajjad, Phys. Rev. C \textbf{76} (2007) 055803.


%\bibitem{SPZ15} M. Spieker, S. Pascu, A. Zilges, and F. Iachello, Phys. Rev. Lett. \textbf{114}, 192504 (2015)

\bibitem{KOJ15} 
M. K. G. Kruse, W. E. Ormand, and C. W. Johnson, arXiv:1502.03464


\bibitem{BG77} P.J. Brussard and P.W.M. Glaudemans, \textit{Shell-model applications 
in nuclear spectroscopy} (North-Holland Publishing Company, Amsterdam 1977). 


\bibitem{BIGSTICK} C. W. Johnson. W. E. Ormand, and P. G. Krastev, Comp. Phys. Comm. \textbf{182} (2013) 2235.



\bibitem{br06} B.A. Brown and W.A. Richter, Phys. Rev. C {\bf74} 
(2006) 034315.

\bibitem{BF66} H. Bannerjee and J. B. French, Phys. Lett. \textbf{23} (1966) 245; J. B. French, Phys. Lett. \textbf{23} (1966) 248.


\bibitem{Fre67} J. B. French, Phys. Lett. \textbf{B 26} (1967) 75.

\bibitem{Fre71}J.B. French and K.F. Ratcliff, Phys. Rev. C {\bf 3} (1971) 94.


\bibitem{Mon75} K. K. Mon and J. B. French, Ann. Phys. {\bf 95} (1975) 90.


\bibitem{Fre83} J.~B.~French and V.~K.~B.~Kota, Phys.~Rev.~Lett.~{\bf 51}
 (1983) 2193; V.~K.~B.~Kota, V.~Potbhare and P.~Shenoy, 
Phys.~Rev.~C~{\bf 34} (1986) 2330.

\bibitem{Won86}  S.~S.~M.~Wong, {\it Nuclear Statistical Spectroscopy},
Oxford Press (New York, 1986).




\bibitem{LSD14}    K.D. Launey,     S. Sarbadhicary,    T. Dytrych,    J.P. Draayer
Comp. Phys. Comm. \textbf{185} (2014) 254.

\bibitem{SHZ13} R. A. Sen'kov, M. Horoi. and V. G. Zelevinksy, Comp.~Phys.~Comm.~\textbf{184} (2013) 215.

\bibitem{JNO01} C. W. Johnson, J. Nabi, and W. E. Ormand, arXiv:nucl-th/0111068.


%\bibitem{BW88}  B. A. Brown and B. H. Wildenthal, Annu. Rev. Nucl. Part. Sci. 38, 29 (1988).

%\bibitem{SMreview05} E. Caurier, G. Martinez-Pinedo, F. Nowacki, A. Poves, and A. P. Zuker, Rev. Mod. Phys. \textbf{77}, 427 (2005). 



%\bibitem{Lanczos} R. R. Whitehead, A. Watt, B. J. Cole, and I. Morrison,  Adv. Nucl. Phys. \textbf{9}, 123 (1977).

%\bibitem{Edmonds} A. R. Edmonds,  \textit{Angular Momentum in Quantum Mechanics},
%(Princeton University Press, Princeton, 1960).


%\bibitem{KB3G} A. Poves, J. S\'anchez-Solano, E. Caurier and F.
 % Nowacki,  Nucl. Phys. A \textbf{694}, 157 {2001}.

\bibitem{CK65}  S. Cohen and D. Kurath, Nucl. Phys. \textbf{73} (1965) 1.

\bibitem{Wildenthal}
B.H. Wildenthal, Prog. Part. Nucl. Phys. \textbf{11} (1984) 5 .



\bibitem{MK75}  D. J. Millener and D. Kurath, Nucl. Phys \textbf{A 255} (1975) 315 .






\end{thebibliography}
\end{document}